# Comment on

# Sound Dispersion in Single-Component System

Boris V. Alexeev

Moscow Academy of Fine Chemical Technology (MITHT)

Prospekt Vernadskogo, 86, Moscow 119570, Russia

B.Alexeev@ru.net

**Abstract**

This paper was initiated by publication of Ref. [1] and can be considered as Comments on Ref. [1]. Authors of Ref. [1] investigate analytically the propagation of sound waves in one component monatomic gas (especially for the intermediate Knudses number region) using methods of local kinetic description in area where they are not applied in principal.

## 1. About basic principles of the Generalized Boltzmann Physical Kinetics and non-local physics.

This paper was initiated by publication of Ref. [1] and can be considered also as Comments on Ref. [1]. Authors of Ref. [1] investigate analytically the propagation of sound waves in one component monatomic gas (especially for the intermediate Knudses number region) using methods of local kinetic description in area where mentioned methods are not applied in principal. Obviously authors [1] don't understand the principles of non-local physics and origin of the Generalized Boltzmann Equation (GBE, known in literature also as Alexeev-Boltzmann equation).

I believe that should be useful to clarify basic principles of non-local physics because application of methods of local physics in situation where local description is not applied in principal is typical in modern publications. For simplicity, I'll consider fundamental methodic aspects from the qualitative standpoint of view avoiding excessively cumbersome formulas. A rigorous description is found, for example, in the monograph [2].

Transport processes in open dissipative systems are considered in physical kinetics. Therefore, the kinetic description is inevitably related to the system diagnostics. Such an element of diagnostics in the case of theoretical description in physical kinetics is the concept of the physically infinitely small volume (**PhSV**). The correlation between theoretical description and system diagnostics is well-known in physics. Suffice it to recall the part played by test charge in electrostatics or by test circuit in the physics of magnetic phenomena. The traditional definition of **PhSV** contains the statement to the effect that the **PhSV** contains a sufficient number of particles for introducing a statistical description; however, at the same time, the **PhSV** is much smaller than the volume $V$ of the physical system under consideration; in a first approximation, this leads to local approach in investigating the transport processes. It is assumed in classical hydrodynamics that local thermodynamic equilibrium is first established within the **PhSV**, and only after that the transition occurs to global thermodynamic equilibrium if it is at all possible for the system under study. Let us consider the hydrodynamic description in more details from this point of view. Assume that we have two neighboring physically infinitely small volumes $PhSV_1$ and $PhSV_2$ in a nonequilibrium system. The one-particle distribution function (DF) $f_{sm,1}(\mathbf{r}_1, \mathbf{v}, t)$ corresponds to the volume **PhSV$_1$**, and the function $f_{sm,2}(\mathbf{r}_2, \mathbf{v}, t)$ — to the volume **PhSV$_2$**. It is assumed in a first approximation that $f_{sm,1}(\mathbf{r}_1, \mathbf{v}, t)$ does not vary within **PhSV$_1$**, same as $f_{sm,2}(\mathbf{r}_2, \mathbf{v}, t)$ does not vary within the neighboring volume **PhSV$_2$**. *It is this assumption of locality that is implicitly contained in the Boltzmann equation (BE).*

However, the assumption is too crude. Indeed, a particle on the boundary between two volumes, which experienced the last collision in **PhSV$_1$** and moves toward **PhSV$_2$**, introduces information about the $f_{sm,1}(\mathbf{r}_1, \mathbf{v}, t)$ into the neighboring volume **PhSV$_2$**. Similarly, a particle on the boundary between two volumes, which experienced the last collision in **PhSV$_2$** and moves toward **PhSV$_1$**, introduces information about the DF $f_{sm,2}(\mathbf{r}_2, \mathbf{v}, t)$ into the neighboring volume **PhSV$_1$**. The relaxation over translational degrees of freedom of particles of like masses occurs during several collisions.

*As a result, "Knudsen layers" are formed on the boundary between neighboring physically infinitely small volumes, the characteristic dimension of which is of the order of path length $l$.*

Therefore, a correction must be introduced into the DF in the **PhSV**, which is proportional to the mean time between collisions and to the substantive derivative of the DF being measured. Rigorous derivation can be found in the book [2]. Very important to notice, that *strict result cannot be obtained as hypothetical limit case from local transport equations, for example from Navier-Stokes equations.*

Let a particle of finite radius be characterized as before by the position **r** at the instant of time *t* of its center of mass moving at velocity **v**. Then, the situation is possible where, at some instant of time *t*, the particle is located on the interface between two volumes. In so doing, the lead effect is possible (say, for **PhSV$_2$**), when the center of mass of particle moving to the neighboring volume **PhSV$_2$** is still in **PhSV$_1$**. However, the delay effect takes place as well, when the center of mass of particle moving to the neighboring volume (say, **PhSV$_2$**) is already located in **PhSV$_2$** but a part of the particle still belongs to **PhSV$_1$**. This entire complex of effects defines non-local effects in space and time.

Non-local effects can be demonstrated with the help of animation – motion of particles – hard spheres in neutral gas. In Moscow Academy of Fine Chemical Technology the animation film (idea belongs to B.V. Alexeev, authors B.V. Alexeev, E. Michailova) is created which demonstrates non-local effects. (This film is demonstrated on the site of CENTRE OF THE THEORETICAL FOUNDATIONS OF NANOTHECHNOLOGY, look at http://www.nonlocalphysics.net.ru ). I invite readers to visit this site where animation film transforms the problem of non-local description into the transparent physical picture without cumbersome formulas.

Let us consider the corresponding animation picture. A volume contains one component rarefied gas in which can be observed predominantly only pair collisions. All nonequilibrium space system is divided on regular open subsystems of physically small volumes (**PhSV**). In a time moment all particles belonging to different **PhSV** are colored in different colors, but have the same color inside of the individual **PhSV**. It is done for the better observation of the following evolution of physical system. Share out one of these **PhSV** (say **PhSV$_1$**). All particles in this **PhSV$_1$** are painted in blue color. The particles are sufficiently numerous for introduction of the local statistical description on the level of one particle distribution function (DF) $f$. Local description (typical for local thermodynamic equilibrium) supposes that all **PhSV$_1$** can be characterized by the same DF (say $f_1$). In neighboring adjacent physically small volumes all particles in the mentioned time moment are painted in other colors which correspond to another distribution functions. The particles from neighboring adjacent physically small volumes are carriers of information about other DF. After penetrating into "blue" volume these carriers are fitting to new DF of neighboring volume after several collisions (for relaxation of particles of the same masses). Let us suppose for simplicity that this process needs only one collision. In another words the "foreign" penetrating particle change its color after the first collision with the "native", blue particle.

You can observe in film as in separated **PhSV₁** ("blue" volume) the boundary many-colored zones are displaying. The character lengths of these zones are of the mean paths *between* collisions.

Obviously this boundary effects can be observed always for elected diagnostic method regardless of special features of separation of the total macroscopic volume $\sim L^3$ by set of infinitely small volumes (**PhSV**).

This consideration clearly demonstrates the existing of non-local effects proportional mean length between collisions $l$, mean time between collisions or Knudsen number. It also means that every physically small volume has boundary Knudsen layers as zones of adjusting of one DF to another DF of neighboring adjacent physically small volume for reduced description on the level of one particle DF.

If the mean path between collision $\lambda$ is of order of the hydrodynamic length $L$ (the more so as the character length of **PhSV₁**) the mentioned Knudsen zones became larger and larger and for intermediate Knudsen numbers ($Kn = l/L \sim 1$) these zones *cover practically the total volume* **PhSV₁**.

*It means that for intermediate Knudsen numbers local Boltzmann kinetic description has no physical sense.*

Let us give the additional explanations following from application of method of many scales to solution of kinetic equation. The physically infinitely small volume (**PhSV**) is an *open thermodynamic system for any division of macroscopic system by a set of PhSVs*. However, the BE

$$Df/Dt = J^B, \qquad (1)$$

where $J^B$ is the Boltzmann collision integral and $D/Dt$ is a substantive derivative, fully ignores non-local effects and contains only the local collision integral $J^B$. The foregoing non-local effects are insignificant only in equilibrium systems, where the kinetic approach changes to methods of statistical mechanics.

This is what the difficulties of classical Boltzmann physical kinetics arise from. Also a weak point of the classical Boltzmann kinetic theory is the treatment of the dynamic properties of interacting particles. On the one hand, as follows from the so-called "physical" derivation of the BE, Boltzmann particles are regarded as material points; on the other hand, the collision integral in the BE leads to the emergence of collision cross sections.

The rigorous approach to derivation of kinetic equation relative to one-particle DF $f$ ($KE_f$) is based on employing the hierarchy of Bogoliubov equations. Generally speaking, the structure of $KE_f$ is as follows:

$$\frac{Df}{Dt} = J^B + J^{nl}, \qquad (2)$$

where $J^{nl}$ is the non-local integral term. An approximation for the second collision integral is suggested by me in *generalized* Boltzmann physical kinetics,

$$J^{nl} = \frac{D}{Dt}\left(\tau \frac{Df}{Dt}\right). \qquad (3)$$

Here, $\tau$ is the mean time *between* collisions of particles, which is related in a hydrodynamic approximation with dynamical viscosity $\mu$ and pressure $p$,

$$\tau\, p = \Pi \mu, \qquad (4)$$

where the factor $\Pi$ is defined by the model of collision of particles: for neutral hard-sphere gas, $\Pi=0.8$. All of the known methods of deriving kinetic equation relative to one-particle DF lead to approximation (3), including the method of many scales, the method of correlation functions, and the iteration method. From Eqs. (2), (3) follow

$$\frac{D}{Dt}\left[f - \tau \frac{Df}{Dt}\right] = J^B. \qquad (5)$$

Note, however, that the second term in Eqn (5) cannot be ignored even for small Knudsen numbers because in that case $Kn$ acts as a small coefficient of higher derivatives, with an unavoidable consequence that the effect of this term will be strong in some regions. The neglect of formally small terms is equivalent, in particular, to dropping the (small-scale) Kolmogorov turbulence from consideration.

Let us consider the proposed by me approximation for the non-local collision integral $J^{nl}$. The structure of dimensionless kinetic equations [2] after application the method of many scales looks as follows:

$$\frac{D_1 \widehat{f}_1^{\,1}}{D\,\widehat{t}_b} + \frac{d_1 \widehat{f}_1^{\,0}}{d\,\widehat{t}_{l,L}} = \widehat{J}^{\,st,0}, \qquad (6)$$

where the second term on the left and first term on the right transform in substantive derivation and local Boltzmann collision integral correspondingly after application of procedure of breaking of the hierarchy of Bogoliubov chain. Lower index "1" in distribution function $\widehat{f}$ corresponds to one particle distribution function. The upper index – to the position of the term in asymptotic series

$$f_s = \sum_{\nu=0}^{\infty} f_s^{\nu}\left(t_b, \mathrm{r}_{ib}, \mathrm{v}_{ib}; t_l, \mathrm{r}_{il}, \mathrm{v}_{il}; t_L, \mathrm{r}_{iL}, \mathrm{v}_{iL}\right)\varepsilon^{\nu}, \qquad (7)$$

where $\varepsilon$ is the density parameter.

Equation (6) contains linking not only with respect to the lower but also with respect to the upper index, implying that in order to employ the kinetic equation, additional assumptions should be made to reduce the equation to one dependent variable.

It is crucial that the term $D_1 \widehat{f}_1^{\,1}/D\widehat{t}_b$ in Eqn (6), reflecting the non-local effects, is of the same order of magnitude as the $l$- and $L$-scale terms. The (unjustified) formal neglect of the term $D_1 \widehat{f}_1^{\,1}/D\widehat{t}_b$ reduces Eqn (6) to the Boltzmann equation.

We intend to employ Eqn (6) for description of evolution of distribution function $\widehat{f}_1^{\,0}$ in $l$- and $L$- scales. But kinetic equation (6) contains linking term $D_1 \widehat{f}_1^{\,1}/D\widehat{t}_b$ with respect to the upper index. The problem arises concerning of approximation this term.

This term allows the exact representation using the series (7).

$$\frac{D_1 \widehat{f}_1^{\,1}}{D\,\widehat{t}_b} = \frac{D_1}{D\,\widehat{t}_b}\left[\frac{\partial \widehat{f}_1}{\partial \varepsilon}\right]_{\varepsilon=0} \qquad (8)$$

Note, however, that in the "field" description the distribution function $f_1$ at the interaction $r_b$-scale depends on $\varepsilon$ through the dynamical variables $\mathbf{r}$, $\mathbf{v}$, $t$, interrelated by the laws of classical mechanics. We can therefore use the approximation

$$\frac{D_1}{D\hat{t}_b}\left[\left(\frac{\partial \hat{f}_1}{\partial \varepsilon}\right)_{\varepsilon=0}\right] \cong \frac{D_1}{D(-\hat{t}_b)}\left[\frac{\partial \hat{f}_1}{\partial (-\hat{t}_b)}\left(\frac{\partial (-\hat{t}_b)}{\partial \varepsilon}\right)_{\varepsilon=0} +\right.$$

$$\left. + \frac{\partial \hat{f}_1}{\partial \hat{\mathbf{r}}_b} \cdot \frac{\partial \hat{\mathbf{r}}_b}{\partial (-\hat{t}_b)}\left(\frac{\partial (-\hat{t}_b)}{\partial \varepsilon}\right)_{\varepsilon=0} + \frac{\partial \hat{f}_1}{\partial \hat{\mathbf{v}}_b} \cdot \frac{\partial \hat{\mathbf{v}}_b}{\partial (-\hat{t}_b)}\left(\frac{\partial (-\hat{t}_b)}{\partial \varepsilon}\right)_{\varepsilon=0}\right] = \quad (9)$$

$$= -\frac{D_1}{D\hat{t}_b}\left[\left(\frac{\partial \hat{t}_b}{\partial \varepsilon}\right)_{\varepsilon=0}\frac{D_1\hat{f}_1}{D\hat{t}_b}\right] \cong -\frac{D_1}{D\hat{t}_b}\left[\left(\frac{\partial \hat{t}_b}{\partial \varepsilon}\right)_{\varepsilon=0}\frac{D_1\hat{f}_1^0}{D\hat{t}_b}\right].$$

The approximation introduced here proceeds against the course of time and corresponds to the condition that there be no correlations as to $t_0 \to -\infty$, where is some instant of time $t_0$ on the $r_b$-scale at which the particles start to interact with each other. The term $\hat{\tau} = \varepsilon\left(\frac{\partial \hat{t}_b}{\partial \varepsilon}\right)_{\varepsilon=0}$ corresponds to mean time $\tau$ between collisions, [2].

The last step in approximation (9) contains the exchange $\hat{f}_1 \leftrightarrow \hat{f}_1^0$. In principal this step can be improved by different ways [2] using series (7). As result we have for example the following kinetic equations

$$\frac{D}{Dt}\left[f_1^0 - \tau\frac{D}{Dt}\left[f_1^0 - \tau\frac{D}{Dt}\left[f_1^0 - \tau\frac{D}{Dt}\left[f_1^0 - ...\right]\right]\right]\right] = J^B, \quad (10)$$

$$\frac{D}{Dt}\left[f_1^0 - \tau\frac{Df_1^0}{Dt} + \tau\frac{D}{Dt}\left(\tau\frac{Df_1^0}{Dt}\right) - \tau\frac{D}{Dt}\left(\tau\frac{D}{Dt}\left(\tau\frac{Df_1^0}{Dt}\right)\right) + ...\right] = J^B. \quad (11)$$

Another types of the non-local collision integral approximation are considered in [2].

## 2. Conclusion.

Now I formulate conclusions of the principal significance:
1. In the Boltzmann local kinetic theory the additional terms of non-local origin are lost. These terms strictly speaking are of the same order as the "classical" ones. From position of non-local physics Boltzmann kinetic theory belongs only to plausible models and don't deliver even minimized model. The GBE introduces a local differential approximation for non-local collision integral.
2. If Knudsen number tends to unity the local approximation of non-local collision integral needs in taking into account the following terms in series containing substantive orders of highest degrees (see the monograph B.V. Alexeev "Generalized Boltzmann Physical Kinetics", Elsevier, 2004). Here we are faced in fact with the "price – quality" problem familiar from economics. That is, what price – in terms of increased complexity of kinetic equation – are we ready to pay for improved quality of the theory? The derived kinetic equations of highest orders have mainly theoretical significance because of its complexity in applications. Moreover the Generalized Boltzmann Equation derived by B.V. Alexeev delivers sufficient approximation for this regime with the correct descriptions of limiting cases. Important, that generalized hydrodynamic equations deliver through approximation in extremely wide diapason of the Knudsen variety sufficient in applications when we enforced to consider much more complicated problems like turbulent flows over blunt bodies. Of course no reason to pay for unreasonable approximations delivered in [1] and applied to BE which is not appropriate in the considered physical situation. In literature you can find papers (typical is Ref. [1]) devoted of application of moment equation of tremendous tensor degrees (especially in the theory of sound propagation for the intermediate Knudsen numbers) for local description. As you see this way has no physical sense.

3. Let us consider some results delivered in [1] from this point of view. It is well known that the break down of the hydrodynamic approach based on Boltzmann physical kinetics, occurs as the frequency of the applied oscillation $v_{osc}$ approaches the collision frequency $v = 1/\tau$ of the gas. Of course no chance to obtain reasonable results for area where $Kn \sim 1$ using local perturbation methods for local kinetic equation like the method of Wang Chang and Uhlenbeck (WCU). Sirovich and Turber were trying to overcome the obvious discrepancies between theoretical and experimental results by introducing in solution of WCU method the additional "empirical" term containing the coefficient $\lambda$ for fitting the theory to experimental data. From position of non-local physics this procedure is the incorrect implicit attempt to obtain approximation for non-local collision integral with $\lambda$ as a parameter of non-locality. Of course this forcing the data to fit experiment leads to crash in area where $1/Kn$ (in [1] Knudsen number is defined as $Kn = v_{osc}/v$) is less than 0.1 and to appearance of the "cutoff" frequencies which have no physical sense. In particular these "cutoff" frequencies are different for different orders of approximations.

4. It seems from the first glance that numerical methods connected with the direct simulation of the particle movement (like DSMC) have no mentioned shortages. But it is not so. Really, these methods unavoidably lead to introduction of physically small volume (**PhSV**) for the moment calculations. These moments are the same in all points of the cell corresponding of the **PhSV**. The numerical fluctuations should be damped by the introduction of filter equations, especially in the case of slow flows by the intermediate Knudsen numbers. It is known from the literature that in many cases the Navies-Stokes equations are taken as the filter equations. This fact returns us to the previous criticism of the Navier-Stokes theory as direct consequence of BE and therefore to inevitable application of the generalized non-local hydrodynamics.

5. Fluctuation effects occur in any open thermodynamic system bounded by a control surface transparent to particles. GBE (2) leads to generalized hydrodynamic equations as the local approximation of non local effects, for example, to the continuity equation

$$\frac{\partial \rho^a}{\partial t} + \frac{\partial}{\partial \mathbf{r}} \cdot (\rho \mathbf{v}_0)^a = 0, \quad (11)$$

where $\rho^a$, $\mathbf{v}_0^a$, $(\rho \mathbf{v}_0)^a$ are calculated in view of non-locality effect in terms of gas density $\rho$, hydrodynamic velocity of flow $\mathbf{v}_0$, and density of momentum flux $\mathbf{v}_0$; for locally Maxwellian distribution, $\rho^a$, $(\rho \mathbf{v}_0)^a$ are defined by the relations

$$(\rho - \rho^a)/\tau = \frac{\partial \rho}{\partial t} + \frac{\partial}{\partial \mathbf{r}} \cdot (\rho \mathbf{v}_0), \ (\rho \mathbf{v}_0 - (\rho \mathbf{v}_0)^a)/\tau = \frac{\partial}{\partial t}(\rho \mathbf{v}_0) + \frac{\partial}{\partial \mathbf{r}} \cdot \rho \mathbf{v}_0 \mathbf{v}_0 + \vec{\vec{I}} \cdot \frac{\partial p}{\partial \mathbf{r}} - \rho \mathbf{a}, \quad (12)$$

where $\vec{\vec{I}}$ is a unit tensor, and $\mathbf{a}$ is the acceleration due to the effect of mass forces.

6. The recommendations of Napier and Shizgal for following development of local analytical methods of the solution of local kinetic equations in the extremely non-local area have no physical sense and are not useful.

In conclusion it should be underlined that in the general case, the parameter $\tau$ is the non-locality parameter; non-local physics lead to new quantum physics [3, 4] which contains

Schrödinger theory as a deep particular case. The violation of Bell's inequalities is found for local statistical theories, and the transition to the non-local description is inevitable [5].

REFERENCES
[1] Napier D.G., Shizgal B.D. Sound Dispersion in Single-Component Systems. Physica A 387, 4099 – 4118, (2008)
[2] Alexeev B.V., Generalized Boltzmann Physical Kinetics, Elsevier, (2004)
[3] Alexeev B.V., J. Nanoelectron. Optoelectron 3, 143 - 158 (2008)
[4] Alexeev B.V., J. Nanoelectron. Optoelectron 3, 316 - 328 (2008)
[5] Bell J.S. Physics 1, 195 (1964)